\documentclass[UKenglish, twocolumn, amsmath, amssymb, prl, aps, floatfix, superscriptaddress, 10pt]{revtex4-2}

\DeclareUnicodeCharacter{0308}{}

\usepackage{listings}
\usepackage{color}
\usepackage{xcolor}
\usepackage{graphicx}
\usepackage{colortbl}
\usepackage{booktabs}
\usepackage{caption}

\usepackage{physics}
\usepackage{braket}
\usepackage{ulem}
\usepackage[T1]{fontenc}

\usepackage[dvipsnames]{xcolor}
\usepackage{tabularray}
\usepackage{hyperref}

\definecolor{tocblue}{RGB}{0,0,120}
\definecolor{figred}{RGB}{200,0,0}

\hypersetup{
  colorlinks=true,
  linkcolor=figred,
  citecolor=green,
  urlcolor=figred
}

\usepackage{orcidlink} % ORCID icon package (RevTeX supports it)
\usepackage{soul}
\usepackage[uncertainty-mode = compact, separate-uncertainty-units = single,  bracket-unit-denominator = false]{siunitx}

\makeatletter

\newcommand{\listsubsectioninsi}{%
  \@starttoc{sitoc}%
}

\makeatother

\AtBeginDocument{\RenewCommandCopy\qty\SI} % avoid conflict with Physics
\usepackage{booktabs} % prettier tables
\usepackage{bm}

\newcommand{\cnn}{Universit\'e Paris-Saclay, Centre de Nanosciences et de Nanotechnologies, CNRS, 10 Boulevard Thomas Gobert, 91120, Palaiseau, France}
\newcommand{\cnnCite}{Universit\'e Paris Cité, Centre de Nanosciences et de Nanotechnologies, CNRS, 10 Boulevard Thomas Gobert, 91120, Palaiseau, France}
\newcommand{\quandela}{Quandela, 7 Rue Léonard de Vinci, 91300 Massy, France}
\newcommand{\nancy}{
Université de Lorraine, CNRS, LPCT, F-54000 Nancy, France}
\newcommand{\majulab}{MajuLab, CNRS-UCA-SU-NUS-NTU International Joint Research Laboratory, 117543 Singapore, Singapore}
\newcommand{\cqt}{Centre for Quantum Technologies, National University of Singapore, 117543 Singapore, Singapore.}
\newcommand{\dtu}{DTU Electro, Department of Electrical and Photonics Engineering, Technical University of Denmark, \O rsteds Plads 343, 2800 Kongens Lyngby, Denmark}
\newcommand{\rome}{Dipartimento di Fisica, Sapienza Università di Roma, Piazzale Aldo Moro 5, 00185 Roma, Italy}
\newcommand{\pdb}{Institute for Photonic Quantum Systems (PhoQS), Center for Optoelectronics and Photonics Paderborn
(CeOPP) and Department of Physics, Paderborn University, Warburger Straße 100, 33098 Paderborn,
Germany}

\begin{document}

\title{Industry-ready spin--photon interfaces for hybrid photonic quantum computing}
\affiliation{\cnn}
\affiliation{\quandela}
\affiliation{\majulab}
\affiliation{\cqt}
\affiliation{\dtu}
\affiliation{\pdb}
\affiliation{\rome}
\affiliation{\cnnCite}
\affiliation{\nancy}
\affiliation{These authors contributed equally to this work.}

\author{H\^elio Huet\,\orcidlink{0000-0002-6267-8752}}
\affiliation{\cnn}
\affiliation{\quandela}
\affiliation{These authors contributed equally to this work.}
\thanks{These authors contributed equally to this work.}

\author{Hubert Lam\,\orcidlink{0000-0002-1712-9201}}
\affiliation{\cnn}
\affiliation{These authors contributed equally to this work.}
\thanks{These authors contributed equally to this work.}

\author{Thibaut Pollet\,\orcidlink{0009-0008-0694-2185}}
\affiliation{\cnn}
\affiliation{These authors contributed equally to this work.}

\thanks{These authors contributed equally to this work.}
\author{Petr Steindl\,\orcidlink{0000-0001-9059-9202}}
\affiliation{\cnn}
\affiliation{These authors contributed equally to this work.}
\thanks{These authors contributed equally to this work.}

\author{Alice Bernard}
\author{Albert Adiyatullin\,\orcidlink{0000-0002-3023-4532}}
\author{Petr Stepanov}
\author{William Hease}
\affiliation{\quandela}
\author{Victor Guilloux\,\orcidlink{0000-0003-3535-9040}}
\affiliation{\cnn}
\author{Nico Margaria\,\orcidlink{0009-0008-8815-320X}}
\author{Joris Verstraten}
\author{Raksha Singla\,\orcidlink{0000-0003-4403-8425}}
\author{Samuel T. Mister}
\author{Anton Pishchagin\,\orcidlink{0000-0002-0965-6394}}
\author{Lara Couronn\'e}
\author{Samuel Huber  }
\author{David Sebastian}
\affiliation{\quandela}
\author{Duc Duy Tran\,\orcidlink{0009-0000-0427-4509} }
\author{Thi Hao Nhi Nguyen}
\author{Thi Phuong Do }
\author{Joseph Sulpizio}
\affiliation{\quandela}
\author{Yann Portella\,\orcidlink{0009-0006-6362-9246}}
\affiliation{\cnn}
\author{Kiarn T. Laverick, \orcidlink{0000-0002-3688-1159}}
\affiliation{\majulab}
\affiliation{\cqt}
\author{Thinhinane Bennour}
\author{Tomas Alexandre De Sousa}
\author{Davide Stefani}
\author{Mathias Pont\,\orcidlink{0000-0002-9623-675X}}
\author{Maxime Descampeaux\,\orcidlink{0000-0001-5112-4204} }
\author{Bianca Scaparra\,\orcidlink{0009-0009-7420-5934}}
\affiliation{\quandela}
\author{Martin A. Jacobsen\,\orcidlink{0000-0001-7534-6135}}
\affiliation{\dtu}
\author{Klaus D. J\"ons\,\orcidlink{0000-0002-5814-7510}}
\affiliation{\pdb}
\author{Rinaldo Trotta\,\orcidlink{0000-0002-9515-6790}}
\affiliation{\rome}
\author{Aristide Lema\^itre\,\orcidlink{0000-0003-1892-9726}}
\affiliation{\cnn}
\author{Martina Morassi\,\orcidlink{0000-0001-6472-2916}}
\affiliation{\cnn}
\author{Olivier Krebs\,\orcidlink{0000-0002-9076-2708}}
\affiliation{\cnn}
\author{Loïc Lanco\,\orcidlink{0000-0003-3607-7539}}
\affiliation{\cnnCite}

\author{Niels Gregersen\,\orcidlink{0000-0002-8252-8989}}
\affiliation{\dtu}
\author{Alexia Auffèves\,\orcidlink{0000-0003-4682-5684}}
\affiliation{\majulab}
\affiliation{\cqt}
\author{Maria Maffei\,\orcidlink{0000-0001-5183-4716}}
\affiliation{\nancy}
\author{Shane Mansfield\,\orcidlink{0000-0002-2231-2422}}
\affiliation{\quandela}
\author{Jean Senellart\,\orcidlink{0000-0001-9153-5745}}
\affiliation{\quandela}
\author{Thomas Volz\,\orcidlink{0000-0001-9850-4992}}
\affiliation{\quandela}
\author{Viviana Villafañe\,\orcidlink{0000-0001-5361-9702}}
\author{Stephen C. Wein\,\orcidlink{0000-0002-4332-4465}}
\affiliation{\quandela}
\author{Dario A. Fioretto\,\orcidlink{0000-0003-3829-4000}}
\affiliation{\cnn} \affiliation{\quandela}
\author{Sebastien Boissier\,\orcidlink{0000-0003-0467-2093}}
\affiliation{\quandela}
\author{Thi Huong Au\,\orcidlink{ 0000-0003-1940-9052}}
\affiliation{\quandela}
\author{Pascale Senellart\,\orcidlink{0000-0002-8727-1086}}
\email{pascale.senellart-mardon@universite-paris-saclay.fr}
\affiliation{\cnn}

\newcommand{\PeS}[1]{{\color{blue}#1}}
\newcommand{\PS}[1]{{\color{orange}#1}}
\newcommand{\RS}[1]{{\color{red}#1}}
\newcommand{\SB}[1]{{\color{teal}#1}}
\newcommand{\HA}[1]{{\color{olive}#1}}
\newcommand{\HH}[1]{{\color{purple}#1}}
\newcommand{\PT}[1]{{\color{cyan}#1}}
\newcommand{\SW}[1]{{\color{magenta}#1}}

\begin{abstract}

\end{abstract}

\flushbottom
\maketitle

\thispagestyle{empty}

{\bf \noindent
Hybrid photonic quantum computers,  {combining stationary matter qubits and flying photonic qubits}, offer an intrinsically networked and resource-efficient route to large-scale, error-corrected quantum computation.
Their core components are cavity-coupled {matter qubits} that act as light--matter interfaces, enabling: high-efficiency on-demand single-photon generation,  stable {near-unity} photon indistinguishability and spin--multi-photon entanglement.
Semiconductor quantum dots in microcavities are a leading platform for realizing such devices. Yet reaching the performance, reproducibility and spin-coherence thresholds for large-scale error correction remains a major challenge requiring industrial fabrication and control.  
Here we report thousands of monolithic semiconductor quantum-dot devices fabricated using a III--V pilot production-line process compatible with large-scale deployment. Systematic control of source parameters yields state-of-the-art efficiency and supports a path to optical losses below fault-tolerance thresholds.
Using field-quadrature state reconstruction as a stringent joint test of efficiency and indistinguishability, we observe near-unity {photon quantum purity} stable over tens of minutes and a record single-photon Wigner-function negativity.
We further demonstrate seven-partite spin--multi-photon entanglement and spin coherence extendable to microsecond timescales in the low-magnetic-field regime.
Finally, photons from distant sources are as indistinguishable as photons emitted successively by a single source.
These results establish {foundry-compatible} III--V quantum dots as a scalable platform for hybrid photonic quantum computing.
}

Photonic quantum computers (PQCs) offer a natural route to large-scale quantum processors with native optical interconnects, foundry-compatible photonic sources, integrated circuits and GHz clock rates \cite{wang2020integrated,bartolucci2023fusion,psiquantum2025manufacturable}.
Yet in fully photonic architectures, both single-photon generation and entangling operations are intrinsically probabilistic, resulting in substantial multiplexing requirements and resource overheads for fault-tolerant operation~\cite{knill2001scheme,browne2005resource,Li_ResourceCostsFaultTolerant_2015,wein_minimizing_2025}.
Hybrid photonic quantum computing (HPQC) addresses this bottleneck by using spin--photon interfaces as deterministic photon sources, local quantum memories and mediators of spin--multi-photon entanglement. In Fig.~\ref{fig:cavity}a, we introduce a HPQC platform where a stationary spin-qubit layer is coupled to a photonic-qubit layer through near-ideal spin–photon interfaces, allowing computation to proceed across both layers. This platform enables several-orders-of-magnitude resource efficiency gains in fusion-based quantum computing schemes~\cite{wein_minimizing_2025,Paesani2026,manohar2026} and tailored spin--optical schemes that provide both low overheads and relaxed component-wise loss requirements ~\cite{gliniasty_spin-optical_2024,dessertaine_enhanced_2026,chan_practical_2025}.
III--V semiconductor quantum dots (QDs) in cavities are among the most advanced light--matter interfaces~\cite{interface1,interface3,Appel_CoherentSpinPhotonInterface_2021,Coste_ProbingDynamicsCoherence_2023a,HoggSpinPhotonCavityWarburton,Mehdi2024}, combining high-efficiency on-demand generation of indistinguishable single photons~\cite{somaschi_near-optimal_2016,Wang2019,Tomm2021,Thomas2021,ding_high-efficiency_2025,Loredo2026DeterministicQD} with spin--multi-photon entanglement~\cite{Lindner_PhotonicClusterState_2009,cogan2023,Su_ContinuousDeterministicAllphotonic_2024,coste_high-rate_2023, huet_deterministic_2025, Appel_EntanglingHoleSpin_2022a,Meng_DeterministicPhotonSource_2024,Meng_TemporalFusionEntangled_2025,Laccotripes_EntangledPhotonSource_2025,Laccotripes_SpinphotonEntanglementDirect_2024}.
However, fault-tolerant HPQC requires a demanding combination of source efficiency, photon indistinguishability, spin coherence and spin--photon entanglement fidelity, consistently achieved across large numbers of devices \cite{gliniasty_spin-optical_2024,dessertaine_enhanced_2026,chan_practical_2025,chen2026fusion}. This demands precise control of the many parameters governing these characteristics, together with fabrication processes capable of delivering them at scale.

Here we report a  pilot production-line process for the fabrication of thousands of monolithic III--V semiconductor QD--cavity {devices} for HPQC.
We demonstrate reproducible control over the cavity and emitter parameters governing photon extraction, indistinguishability, and multi-device compatibility.
Representative devices exhibit state-of-the-art photon-generation efficiency and near-ideal quantum interference with a reference laser, evidencing photon indistinguishability that remains stable over tens of minutes. Combined with high source efficiency, this enables quadrature-basis state reconstruction and reveals a record single-photon Wigner-function negativity of $-0.330(9)$ at the device output.
Spin-enabled devices generate up to seven entangled qubits on demand, with hole-spin coherence extendable to the microsecond regime, a timescale relevant for error-correction~\cite{gliniasty_spin-optical_2024,dessertaine_enhanced_2026}.
Finally, we show that photons emitted by independent devices interfere at the limit set by the coherence of the individual sources, without filtering or post-selection.
Together, these results establish {foundry-compatible} III--V quantum dots as a scalable spin--photon-interface platform, {with a clear and actionable route toward operating below the thresholds required for fault-tolerant HPQC}.

%=================================================
\section*{An industrial pilot line for semiconductor photon sources }
%=================================================

\begin{figure*}[t]
    \centering
    
    \includegraphics[width=0.8\linewidth]{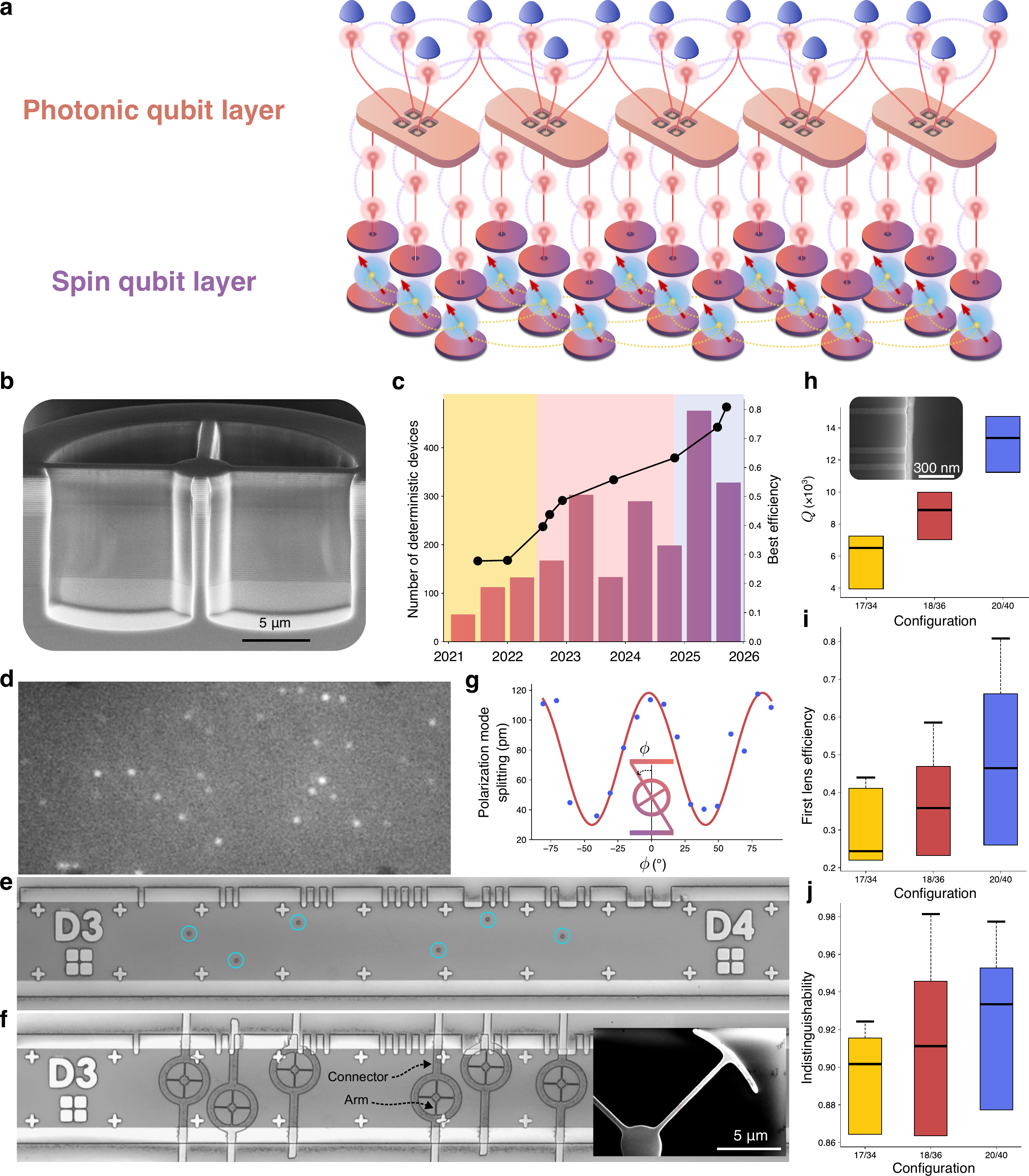}
    \caption{\textbf{Pilot production-line for semiconductor spin-photon interfaces.} 
    \textbf{a,}~Schematic of hybrid photonic quantum computing, with a stationary spin-qubit layer mapped to a photonic-qubit layer through optimal spin–photon interfaces. Photon interference, photon detection and short-term spin memories enable non-local operations and computation across both layers.
    \textbf{b,}~Scanning electron microscopy (SEM) image of a QD-cavity device.
    \textbf{c,}~Number of fabricated single-photon source devices (left axis) and highest first-lens efficiency achieved (right axis) as a function of time. The background colors indicate three generations of production wafers with $n_\text{bottom}=2n_\text{top}$ and $n_\text{top}=17,18,20$.
    \textbf{d,}~Photoluminescence image recorded at \SI{4}{\kelvin} of a sample under wide-field \SI{850}{\nano\meter} excitation showing multiple QDs and alignment markers for QD localization (purple markers) and navigation.
    \textbf{e,}~Optical microscope image acquired after development of the cryogenic resist exposure, showing the resist opening overlaid on one QD in each write field (blue circle).
    \textbf{f,}~Optical microscope image of the same sample after room-temperature exposure and development of the cavity shapes aligned on the previous openings. The inset shows the smallest feature size achieved (\SI{350}{\nano\meter}), demonstrating the high precision of the fabrication process.
    \textbf{g,}~{Fundamental cavity-mode} splitting as a function of the arm and connector angle with respect to the crystalline axis (see inset).
    \textbf{h,}~Measured mean (black bar), standard deviation (coloured box), and best value of the quality factor $Q$ for the three wafer generations. Insert: SEM image of latest generation devices, showing etching verticality of $0\pm0.5^\circ$ degrees and corrugation below $10\pm5\ $\,nm. \textbf{i,j,} Measured mean, standard deviation, and best value of the first lens efficiency and single-photon indistinguishability $M_s$ for the three wafer generations.}
    \label{fig:cavity}
\end{figure*}

Our semiconductor single-photon sources consist of individual InGaAs QDs positioned at the field maximum of high-quality factor (Q)  micropillar cavities~\cite{somaschi_near-optimal_2016}. The devices are fabricated from planar cavity heterostructures comprising AlGaAs/GaAs distributed Bragg reflectors (DBRs), with the bottom mirror containing twice as many layer pairs as the top mirror ($n_{\mathrm{bottom}}=2n_{\mathrm{top}}$).  The structures are doped to form PIN diodes enabling QD charge control and emission-wavelength tunability. Figure~\ref{fig:cavity}b shows a typical pillar device, in which the QD is positioned at the center of the cavity mode. The pillar is connected to a circular frame by narrow ridges (“arms”), and the frame to larger wire-bonding structures by additional ridges (“connectors”). To control the QD-cavity coupling, the positions of the QDs are determined by measuring the spatial dependence of their emission in a two-step process. First, a far-field image with a \SI{850}{\nano\meter} illumination is used to coarsely locate the QDs with respect to fiducial markers.
Each QD is then placed under a focused laser beam and the sample position is scanned to maximize the emission intensity, allowing the QD position to be determined with an accuracy of approximately \SI{50}{\nano\meter}~\cite{dousse_controlled_2008} (Fig.~\ref{fig:cavity}c). A green laser{, aligned with the excitation laser,} then exposes a disk centered on each targeted QD in a resist, which is then developed (Fig.~\ref{fig:cavity}d). The {overlaid openings} serve as marks for a second lithography step performed at room-temperature to define the fully-connected pillar shapes (Fig.~\ref{fig:cavity}e). The diameter of each pillar is adjusted to tune the cavity-mode energy into resonance with the selected QD transition.

Figure~\ref{fig:cavity}f. shows the evolution of device production over the past five years. A similarly large number of additional test devices was also fabricated to systematically investigate the parameters affecting source performance (see Fig.~S8 in SI). 
%\ref{fig:production} in SI).
Device fabrication was first carried out in an academic facility during the development phase before being transferred to a pilot-line production process in mid-2023. The right axis shows the highest device efficiency achieved over time, {defined as the probability of collecting a photon per pulse} {at the first lens}. This metric increased from less than 30\% to 80\% by the end of 2025, {which establishes} a new state-of-the-art for monolithic semiconductor devices. 

This systematic increase in source efficiency has been enabled by full control of all device parameters. The source efficiency is given by $B=p_\mathrm{e}\eta_{\mathrm{Coll}}$, where $p_\mathrm{e}$ is the probability of preparing the QD in its excited state and $\eta_{\mathrm{Coll}}$ is the photon collection efficiency at the device output. In the weak-coupling regime, $\eta_{\mathrm{Coll}}=\eta_{\mathrm{Extract}}\eta_{\mathrm{Cav}}$, where $\eta_{\mathrm{Cav}}=\Gamma_\mathrm{Cav}/(\Gamma_\mathrm{Cav}+\Gamma_\mathrm{Bkg})$ is the probability of emission into the cavity mode and $\eta_{\mathrm{Extract}}=1-Q/Q_\mathrm{Loss}$ is the probability that the photon exits the cavity through the top mirror. Here, $\Gamma_\mathrm{Cav}$ is the decay rate into the cavity mode, controlled by the Purcell factor $F_\mathrm{P}\propto Q/V$, with $Q$ the cavity quality factor and $V$ the mode volume, while $\Gamma_\mathrm{Bkg}$ accounts for QD emission into all other background modes. $Q_\mathrm{Loss}$ accounts for all uncontrolled cavity losses, including bottom-mirror leakage, sidewall scattering and absorption.

The number of top DBR pairs, $n_{\mathrm{top}}$, has been progressively increased from 17 to 20, while the etching recipe was optimized to reduce fabrication imperfections and increase $Q_\mathrm{Loss}$. Numerical simulations show that the etching verticality and DBR sidewall corrugation are important factors limiting source performance (see Supplementary Information). Figure~\ref{fig:cavity}b. and the insert in Fig.~\ref{fig:cavity}h. show scanning electron microscope (SEM) images of optimized micropillars, demonstrating nearly vertical sidewalls and a corrugation depth below \SI{10}{nm}. The fabrication process now routinely yields a sidewall angle of $\SI{0\pm0.3}{\degree}$ and a corrugation of $\SI{30\pm5}{\nano\meter}$ (see Supplementary Information).
Figure~\ref{fig:cavity}h shows the mean, standard deviation and best value of $Q$ for the three generations of devices with increasing $n_{\mathrm{top}}$, illustrating the continuous improvement in fabrication quality. The sources are operated either under resonant excitation (RF) or near-resonant longitudinal acoustic-phonon-assisted excitation (LA)~\cite{LA1,LA2,LA3}, the former yielding near-unity $p_\mathrm{e}$ and the latter {enabling polarization-encoded spin-multiphoton} entanglement generation~\cite{coste_high-rate_2023}.
Fig.~\ref{fig:cavity}i,j present the mean value, standard deviation and best value of the first-lens efficiency and the indistinguishability $M_{s}$ of successively emitted photons for each generation. They reveal a continuous increase in both metrics, with mean (maximal) values of indistinguishability reaching 95\% (98\%) without any spectral filtering to remove phonon-sideband emission. These values already approach and exceed the thresholds required for error correction for photons successively emitted by the same source~\cite{gliniasty_spin-optical_2024}. We demonstrate hereafter that the same performance can be achieved for remote sources.

Although corrugation has been minimized and near-perfect etching verticality achieved, the simulated maximum efficiency of 95\% has not yet been reached. We attribute the remaining losses to {absorption due} to the current doping profile in the Bragg mirrors, which can be reduced through improved doping designs. Finally, we are exploring new cavity designs to further increase the maximum accessible efficiency, as required to fully enter the large-scale error-correction regime beyond threshold. We present a next-generation micropillar design featuring concentric rings and engineered DBRs to further suppress the QD emission toward the sidewalls and substrate, providing an additional 3\% efficiency gain (see Supplementary Information).

\begin{figure*}[t]
    \centering
     \includegraphics[width=0.8\linewidth]{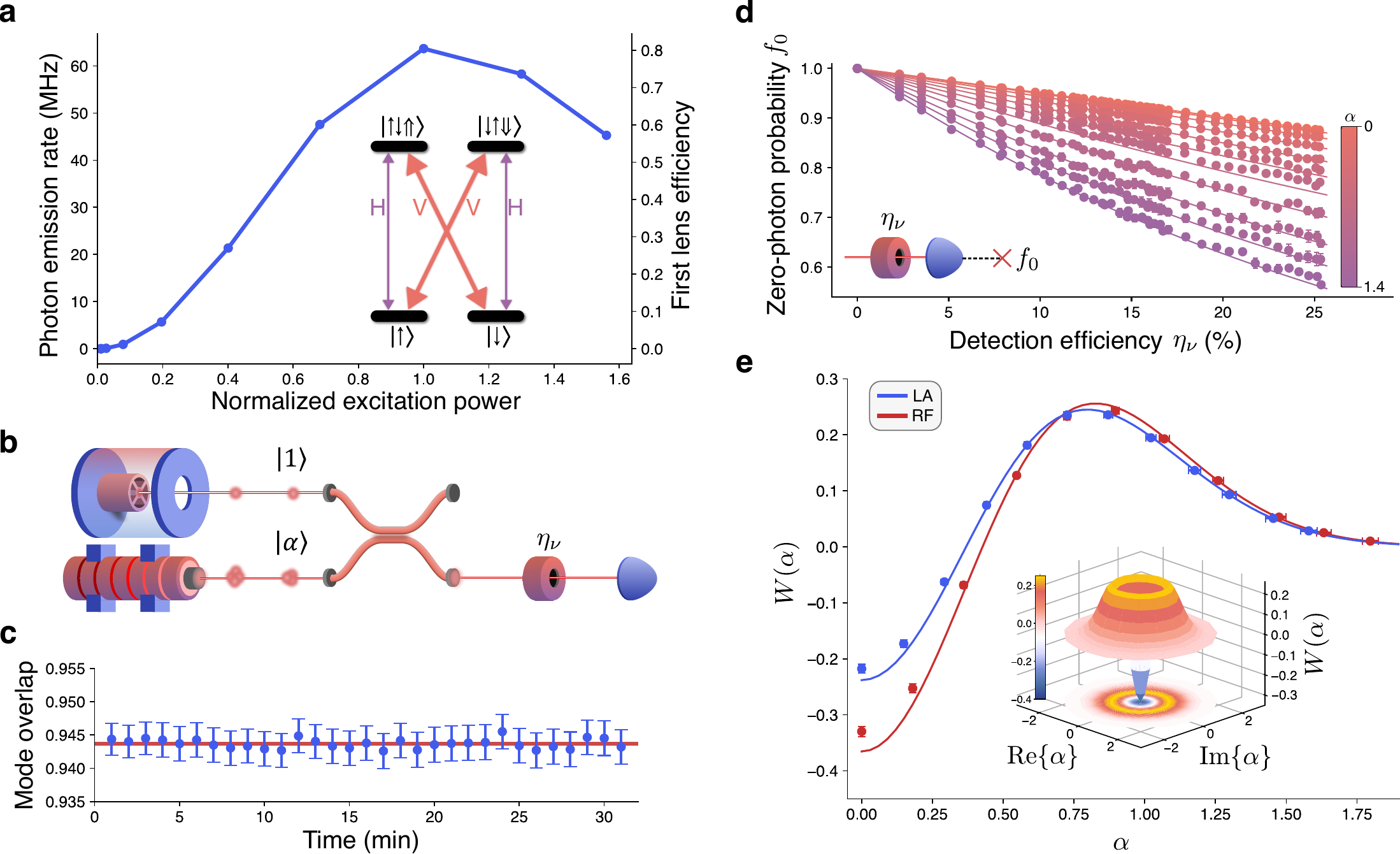}
    \caption{\textbf{Wigner function reconstruction {of single-photon state}.} \textbf{a,}  
    Photon rates at the device output (first lens) under RF excitation. {Inset, optical transitions of the negatively charged QD coupled to a polarized split cavity.} \textbf{b,}~Schematic of the experimental setup used to measure the Wigner function where the photons emitted by the QD are interfered with a laser field (local oscillator) on a beam splitter. The counts are the output are recorded for various controlled attenuations. \textbf{c,}~Measured mode overlap $M$ between the LO and single photons as a function of the measurement time for LA-phonon-assisted excitation. Each data point {represents average value acquired over one minute}. \textbf{d,}~Zero-photon probability $f_0$ (symbols) measured for a set of detection efficiencies $\eta_\nu$ for different phase-space coordinates $\alpha=0-1.8$. The lines correspond to a fit to the experimental data, {enabling the reconstruction of the photon-number distribution of the state $\hat{\rho}$ displaced by $\alpha$.} \textbf{e,} Cross-sections of the reconstructed Wigner functions {of photon emitted} under RF and LA excitation at the device output. The negativity is larger under RF excitation due to the highest source efficiency. Inset, reconstructed Wigner function {of photon produced} under RF excitation.}
    \label{fig:wigner}
\end{figure*}

Depending on the adopted quantum-computing scheme, the same device architecture can be operated either as a polarized single-photon source or as a polarization-encoded spin--multi-photon entanglement generator by tuning the cavity birefringence to achieve a polarized or unpolarized Purcell effect. Here, we set the cavity birefringence through in-plane strain by varying the orientation of the arms and connectors linking the pillars to the large mesas supporting the contacts, as shown in Fig.~\ref{fig:cavity}g. The cavity-mode splitting increases from \qty{30} {\pico\meter} to \qty{150} {\pico\meter} as the structure is rotated from $\theta=45^\circ$ to $\theta=0^\circ$ with respect to the $[110]$ or $[1\bar{1}0]$ crystalline axes. This feature is exploited in Fig.~\ref{fig:wigner} for single-photon generation and in Fig.~\ref{fig:cluster} for spin--multi-photon entanglement.

%=================================================
\section*{Wigner function measurement}
%=================================================

We benchmark our QD single-photon sources through a full reconstruction of the emitted quantum state in the quadrature phase space~{\cite{Wigner1932, Hillery1984, Lvovsky2009_review}}. This approach goes beyond the standard test of photon indistinguishability based on the Hong-Ou-Mandel interference, which probes {the quantum purity of the generated photon states on timescales of up to a few microseconds~\cite{Tomm2021}}. Reconstructing the Wigner function requires  a near-unity and stable mode overlap between the emitted photons and a reference local oscillator over timescales of minutes. We use a homodyne detection scheme, sketched in Fig.~\ref{fig:wigner}b, in which the single-photon state interferes with a weak local oscillator (LO). This interference implements a  {tunable} phase-space displacement operation $\hat{D}(\alpha)$ required to reconstruct the state over the full quadrature space. {The measurement is performed on} a bright source {for which the} QD hosts a ground-state electron and presents four degenerate linearly polarized transitions (Fig.~\ref{fig:wigner}a inset). By exploiting a polarized Purcell enhancement, we obtain bright V-polarized single-photon emission with a first-lens efficiency of 78\,\% (69\,\%) under RF (LA) excitation, as shown in Fig.~\ref{fig:wigner}a.  The overlap between the single-photon and LO modes is optimized by spectrally shaping the LO laser~\cite{lam_optimizing_2025}. Figure~\ref{fig:wigner}c presents a measured mode overlap of $M=0.944(2)$ between the two fields, stable over thirty minutes, highlighting the exceptionally low-noise solid-state environment surrounding our engineered artificial atoms.

The value of the Wigner function at phase-space coordinate $\alpha$ is given by \mbox{$W(\alpha)=\tfrac{2}{\pi}\Tr{\hat{\Pi}\hat{\rho}(\alpha)}$}, {with the photon-number parity \mbox{$\hat{\Pi}=(-1)^{\hat{n}}$}} of the displaced quantum state \mbox{$\hat{\rho}(\alpha)=\hat{D}(\alpha)\hat{\rho}\hat{D}^\dagger(\alpha)$}~\cite{Royer1985,Banaszek1996, Lvovsky2009_review}, and can be measured from the photon-number distribution of $\hat{\rho}(\alpha)$~\cite{bertet_direct_2002,laiho_direct_2009,laiho_probing_2010}. Owing to the ultrafast ($\approx 100$\,ps) single-photon wavepackets, we access this distribution through multi-photon correlations using measurements of the zero-photon probability $f_0(\eta_\nu,\alpha)$~\cite{zambra_experimental_2005,Rossi2004,allevi_state_2009,Zhang2009}.

We use a single SNSPD as a bucket detector, registering the presence of at least one incident photon without photon-number resolution, after a calibrated attenuator that provides a set of detection efficiencies $\eta_\nu$ (with $\nu=1 \mathrm{~to~}42$). The zero-photon probability $f_0(\eta_\nu,\alpha)$ links the detected click probability $f_\mathrm{det}(\eta_\nu,\alpha)$ to the photon-number distribution of $\hat{\rho}(\alpha)$:
\begin{align}
f_0(\eta_\nu,\alpha)=\sum_{n=0}^{N}(1-\eta_\nu)^n \langle n|\hat{\rho}(\alpha)|n\rangle=1-f_\mathrm{det}(\eta_\nu,\alpha).
\end{align}
The linearity of this relation enables the reconstruction of the full photon-number distribution of $\hat{\rho}(\alpha)$ up to $N$ photons from $N_\nu\ge N$ measurements performed at ideally equidistant $\nu$ detection efficiencies $\eta_\nu$. Figure~\ref{fig:wigner}(d) presents $f_0(\eta_\nu,\alpha)$ for $462$ values of $\eta_\nu$ and $\alpha$. For $\alpha=0$ (top line in Fig.~\ref{fig:wigner}(d)), $f_0(\eta_\nu,0)$ exhibits a linear dependence, whereas the nonlinear behavior of the remaining curves reflects the multi-photon character for $\alpha>0$, enabling reconstruction of the coherences and two-photon component of $\hat{\rho}$. The Wigner function is then corrected for the losses between the interference beamsplitter and the device output (Supplementary Information)~\cite{Leonhardt1997,Lvovsky2009_review, laiho_direct_2009}.

Fig.~\ref{fig:wigner}e presents the profile of the Wigner function along one quadrature for both LA-phonon assisted excitation and RF excitation, at the output of the device, the inset presents the reconstructed Wigner function under RF excitation. We observe negativies {$\mathrm{min}(W(\alpha))$} of \SI{-0.218\pm0.008} and \SI{-0.330\pm0.009}, respectively. The experimental data (symbols) in Fig.~\ref{fig:wigner}e are in excellent agreement with the expected ideal Wigner function for the corresponding source efficiency (lines, see supplementary Information). To the best of our knowledge, this is the first reconstruction of the Wigner function of optical quantum light generated by solid-state emitters. Such achievement evidences the extreme stability of the solid-state environment surrounding our quantum emitter. The observed negativity, directly related to the source efficiency, is larger than the record values reported for on-demand atom-based single-photon sources~\cite{magro_deterministic_2023}.

%=================================================
\section*{Spin--multi-photon entanglement}
%=================================================

\begin{figure*}[t]
    \centering
    \includegraphics[width=0.9\linewidth]{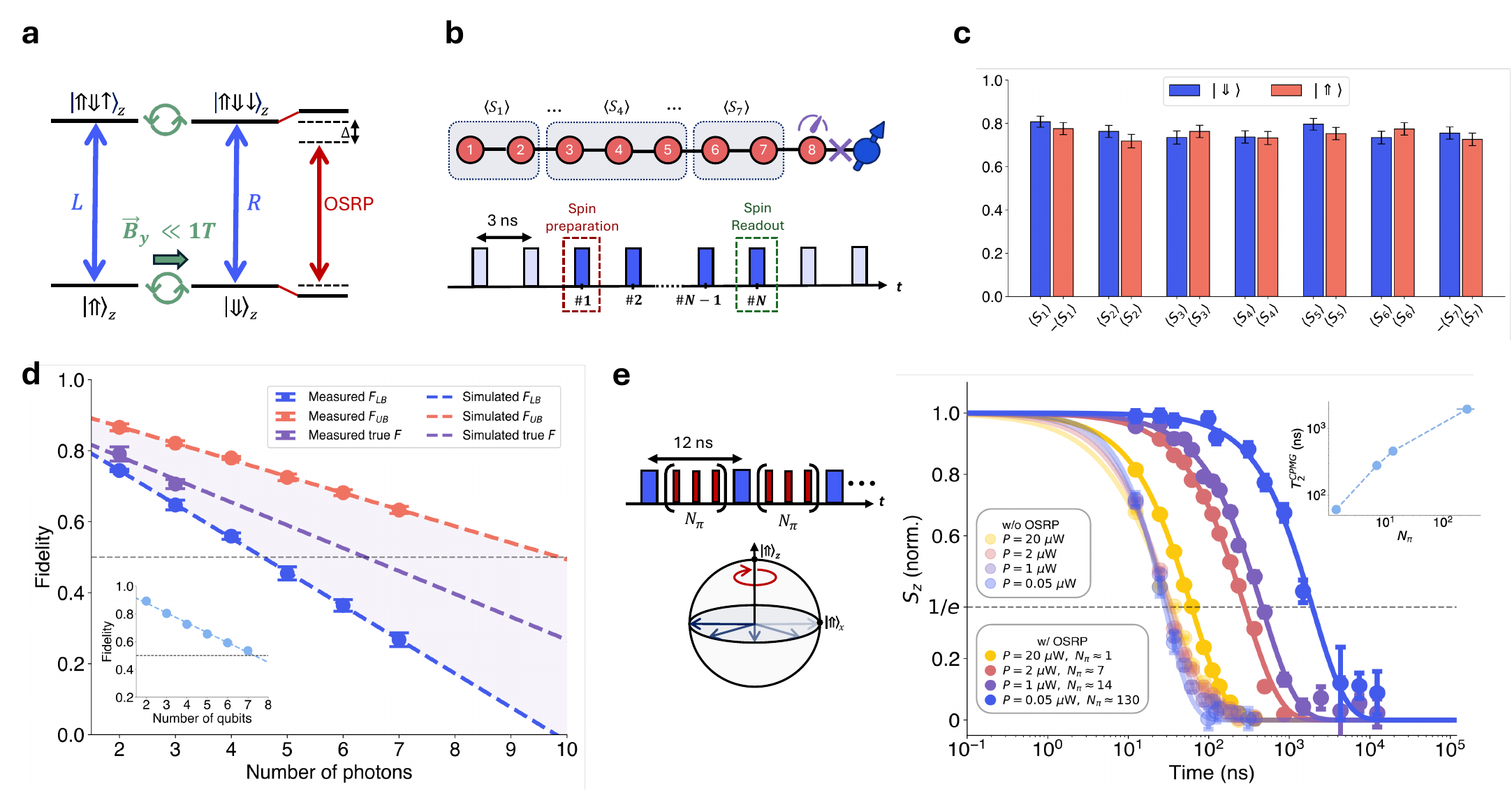}
    \caption{ 
    \textbf{Spin--multi-photon entanglement and dynamical decoupling.} \textbf{a,} Optical selection rules for a positively-charged QD under a weak transverse magnetic field. \textbf{b,} Excitation sequence used to generate spin--multi-photon linear cluster states of arbitrary length. Stabilizers $S_k$ are measured along the photonic chain, highlighted by dotted boxes, for state characterization. Detection of the first photon heralds preparation of the spin state (red box), whereas detection of the final emitted photon implements spin readout (green box), disentangling the spin from the photonic chain (purple cross). \textbf{c,} Measured stabilizer expectation values $\langle S_k \rangle$ for 7-photon linear cluster states with the spin heralded in $\ket{\Downarrow}$ (blue) or $\ket{\Uparrow}$ (orange). \textbf{d,} Multipartite fidelity lower bound (blue), upper bound (orange) and true fidelity (purple) versus number of photons for purely photonic linear cluster states. Each fidelity is averaged over the two state configurations of \textbf{c}. Dashed lines show predictions of the simulation model. Inset, multipartite fidelity versus number of qubits for hybrid spin--multi-photon linear cluster states, inferred from experimental results using the simulation model. \textbf{e,} Left, Dynamical decoupling sequence used to extend spin coherence. Optical rotation pulses (red rectangles) periodically rotate the spin about the optical $z$-axis. The average number of rotation pulses, $N_\pi$, is controlled through the excitation-laser power, which reduces the photon-emission probability. Right, Normalized spin projection $S_z$ versus time for four excitation-laser power values {and corresponding $N_\pi$}. Opaque symbols correspond to measurements with OSRPs applied between excitation pulses and semi-transparent symbols to control measurements without OSRPs. Solid lines are stretched-exponential fits. Inset, scaling of $T_2^{\mathrm{CPMG}}$ versus $N_\pi$.}
    \label{fig:cluster}
\end{figure*}

We now use a QD device hosting a single hole spin with unpolarized Purcell enhancement {and first-lens efficiency of 38\%} to generate spin–multi-photon linear cluster states. The optical selection rules shown in Fig.~\ref{fig:cluster}a directly map the ground spin states $\ket{\Uparrow}$ and $\ket{\Downarrow}$, onto the circular polarizations of the emitted photons, $\ket{R}$ and $\ket{L}$.
The experimental protocol, based on the proposal of Lindner and Rudolph~\cite{Lindner_PhotonicClusterState_2009}, is shown in Fig.~\ref{fig:cluster}b. A periodic train of {linearly-polarized} excitation pulses separated by \qty{3}{\nano\second} triggers single-photon emission from the QD. The delay is chosen such that the spin undergoes a $3\pi/2$ precession between successive emissions under a transverse magnetic field $B=\qty{150}{\milli\tesla}$ applied along the $y$-axis, thereby implementing $R_y(3\pi/2)$ spin gates. Detection of the first photon in the right- (left-) circular polarization basis $R$ ($L$) prepares the spin in $\ket{\Uparrow}$ ($\ket{\Downarrow}$). After precessing into a superposition, the spin becomes entangled with the next emitted photon. Repeating this emission–precession cycle $N-1$ times deterministically generates a linear cluster state comprising the hole-spin qubit and $N-1$ photons. Measuring a final photon in the circular polarization basis disentangles the spin, yielding a purely photonic linear cluster state of $N-2$ photons~\cite{huet_deterministic_2025}.

To certify genuine multipartite entanglement, we evaluate the fidelity $F$ of the generated state with respect to an ideal cluster state using the stabilizer formalism~\cite{Toth_EntanglementDetectionStabilizer_2005}. The stabilizer operators of a linear cluster are \mbox{\(S_k = Z_{k-1} X_k Z_{k+1}\)}, with boundary conditions \mbox{\(S_0=S_{N+1}=I\)}, where $k\in[1,N]$ denotes the photon index. Each stabilizer expectation value is extracted from polarization-resolved measurements on three consecutive photons, $k-1$, $k$ and $k+1$, while photons $k=1$ and $k=N$ are measured in the circular polarization basis for spin preparation and readout, effectively requiring five-photon correlations. Figure~\ref{fig:cluster}c shows the measured stabilizer expectation values for a 7-photon linear cluster obtained from the underlying 9-qubit spin-photon state, yielding an average value of $\langle S_k\rangle=$ \SI{0.76(1)}{}. 

From these measurements, we compute fidelity lower and upper bounds, $F_{LB}$ and $F_{UB}$ (see Supplementary Information). As shown in Fig.~\ref{fig:cluster}d, $F_{LB}>0.5$ is observed for up to four photons. To obtain a more accurate estimate of the true fidelity, which is expected to largely exceed $F_{LB}$ as N increases~\cite{Toth_EntanglementDetectionStabilizer_2005}, we use a numerical model of the measurement process based on independently characterized system parameters (see Supplementary Information). As an independent validation, we directly measure the true state fidelity for photonic cluster states comprising two and three photons, obtaining \SI{0.79(2)} and \SI{0.71(1)}, respectively (purple squares in Fig.~\ref{fig:cluster}d), in excellent agreement with the simulated values of 0.80 and 0.72. The model reproduces the experimental data with an absolute deviation below $1\%$ (dashed lines in Fig.~\ref{fig:cluster}d) {and yields a true state fidelity remaining} above $50\%$ for photonic cluster states comprising up to six photons, {tripling the number of entangled photon compared to previous} deterministic photonic cluster state demonstrations based on QDs~\cite{Su_ContinuousDeterministicAllphotonic_2024,huet_deterministic_2025}. The model further demonstrates that the underlying spin--multi-photon cluster state before spin disentanglement reaches seven genuinely entangled qubits (inset of Fig.~\ref{fig:cluster}d), {with a per-cycle spin-photon entangling fidelity of 93.5(3)\%}. 

Importantly, the cluster-state results presented above are obtained without any mitigation of decoherence induced by the nuclear spin bath, one of the dominant sources of error limiting spin--multi-photon entanglement fidelity~\cite{Urbaszek_NuclearSpinPhysics_2013a,Bechtold_ThreestageDecoherenceDynamics_2015a}. In our devices, this interaction limits the hole-spin coherence time to $\sim\qty{20}{\nano\second}$~\cite{Coste_ProbingDynamicsCoherence_2023a} (Supplementary %Fig.~\ref{fig:SI_fig_spin_merit}a).
Fig.~S13a).
{We therefore address this limiting mechanism directly by implementing dynamical decoupling under the low magnetic-field conditions used for cluster-state generation (100 mT).}

Detuned optical spin-rotation pulses (OSRPs), {interleaved between LA-phonon-assisted excitation pulses}, periodically perform $\pi$ rotations about the optical axis~\cite{Press_UltrafastOpticalSpin_2010,huet_deterministic_2025} (Fig.~\ref{fig:cluster}e). Rather than increasing the delay between excitation and readout through pulse picking~\cite{Zaporski_IdealRefocusingOptically_2023}, we reduce the excitation power, thereby increasing both the average time before readout and the average number of refocusing pulses $N_\pi$ experienced by the spin (see Supplementary material). This implements a Carr-Purcell-Meiboom-Gill (CPMG)-like dynamical decoupling sequence~\cite{Carr1954,Cywinski_HowEnhanceDephasing_2008}. The normalized spin projection along the optical axis, $Sz$, measured at discrete time delays separated by 12 ns and shown in Fig.~\ref{fig:cluster}e, reveals a two-orders-of-magnitude increase in coherence time with increasing $N_\pi$ compared with measurements performed without dynamical decoupling, reaching $\qty{1.94(17)}{\micro\second}$.

%================================================\section*{Identical photons from distant sources}
%=================================================
\begin{figure*}[t]
    \centering
    \includegraphics[width=0.78\linewidth]{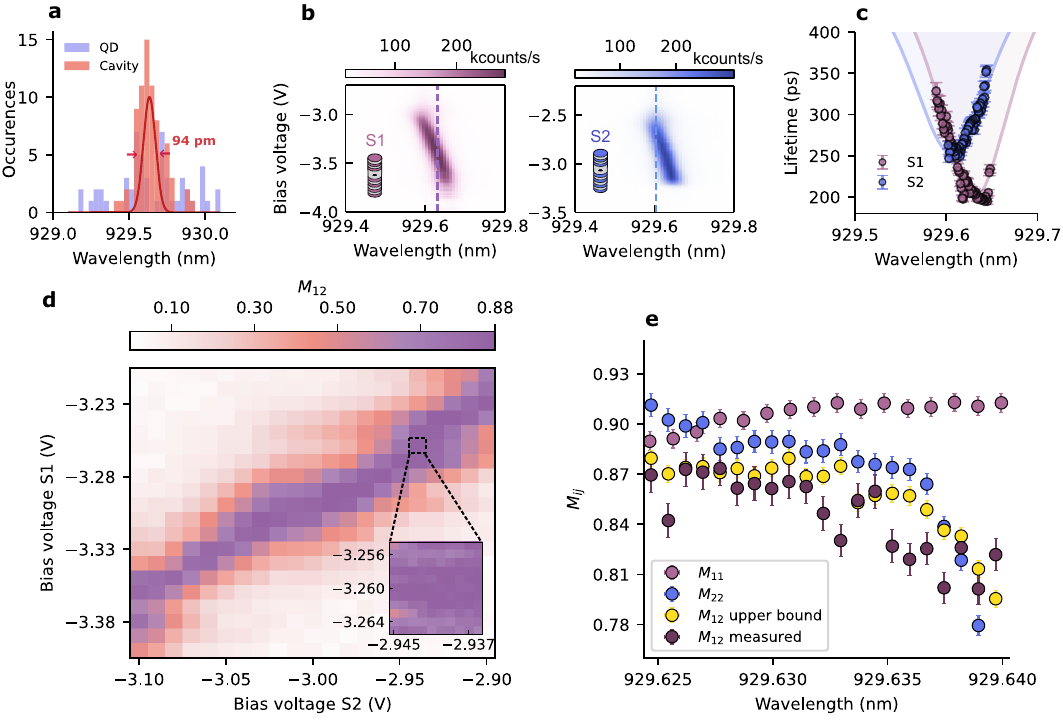}
    \caption{\textbf{Identical photons from distant sources.} \textbf{a,} Wavelength distribution of fabricated micropillar cavities. A normal-distribution fit yields a full width at half maximum of \qty{95(1)}{\pico\meter}. The embedded QDs exhibit a broader wavelength distribution, requiring post-fabrication fine tuning to bring each QD into resonance with its cavity (Supplementary Information). \textbf{b,} Charge plateaus of the negatively charged QD transition for two QDs embedded in separate micropillars, S1 and S2. Dashed lines indicate the cavity resonances. \textbf{c,} Radiative lifetimes of S1 and S2 measured as a function of wavelength by varying the p-i-n bias voltage. Solid lines are Lorentzian fits describing the detuning-dependent Purcell effect. \textbf{d,} Remote-source indistinguishability measured as a function of the bias voltages applied to S1 and S2. At each operating point, a photon-correlation histogram between photons emitted by the two sources is used to extract the mutual indistinguishability (Supplementary Information). The inset shows a finer scan around the maximum HOM visibility. \textbf{e,} Mutual indistinguishability measured as a function of emission wavelength in the absence of spectral detuning between S1 and S2. Experimental values are compared with the theoretical upper bound calculated from the individual indistinguishabilities $M_{11}$ and $M_{22}$ and the corresponding source lifetimes.}
    \label{fig:two_sources}
\end{figure*}

\section*{Identical photons from distant sources}

We now address a requirement for scalability: the ability to generate indistinguishable photons from independent devices~\cite{zhai_quantum_2022}. Because the photon wavepacket is strongly shaped by the QD-cavity interaction through the Purcell effect, this requires tight control of both cavity and emitter properties. Through deterministic fabrication and precise control of the multi-step process, we obtain nominally identical micropillar cavities, with resonance wavelengths distributed according to a Gaussian of FWHM \qty{95(1)}{\pico\meter}, which is narrower than the average cavity linewidth of \qty{116}{\pico\meter} (Fig.~\ref{fig:two_sources}a). The devices feature 18 DBR top pairs and are expected to achieve a photon indistinguishability of \qty{91}{\percent} on average (Fig.~\ref{fig:cavity}j).

We select two devices containing negatively charged QDs (Fig.~\ref{fig:two_sources}b), denoted S1 and S2. Under LA-phonon-assisted excitation, they exhibit unpolarized first-lens efficiencies of \qty{55}{\percent} and \qty{31.2}{\percent}, respectively. Electrical tuning of the QD transitions across the cavity resonances controls the photon wavepackets via the Purcell effect, which strongly reduces the radiative lifetime at resonance (Fig.~\ref{fig:two_sources}c). To maximize the two-source indistinguishability, the emission wavelengths and radiative lifetimes are matched while maintaining both emitters in stable charge states by Coulomb blockade.

To explore the interplay between cavity effects and charge-state stability, we scan the bias voltages of S1 and S2 using an iterative grid-refinement strategy with step sizes of \qty{10}{\milli\volt} and \qty{1}{\milli\volt}, respectively (Fig.~\ref{fig:two_sources}d). At each operating point, an intensity-correlation histogram is acquired to determine the mutual indistinguishability $M_{12}$ (Supplementary Information). No narrow spectral filtering to remove the phonon sideband contribution~\cite{zhai_quantum_2022}, temporal post-selection or active  stabilization is employed. In the absence of emitter detuning, high mutual indistinguishability is maintained across the charge plateaus over a wavelength range of \qty{15}{pm}. The highest HOM visibility, $V_\mathrm{rem}=\qty{85(1)}{\percent}$, corresponds to a mutual indistinguishability of $M_{12}=\qty{88(1)}{\percent}$ after correction for the finite second-order intensity auto-correlation function $g^{(2)}(0)$ of the individual sources (Supplementary Information). Figure~\ref{fig:two_sources}e compares the measured values with the corresponding theoretical upper bounds deduced from the individual source indistinguishabilities $M_{11}$ and $M_{22}$. The measured $M_{12}$ is very close to the upper bound across the full tuning range, demonstrating that uncorrelated noise has a negligible impact on the two-source interference, which is primarily limited by the indistinguishability of consecutively emitted photons from each source. Similar performance is obtained using a third device, yielding $M_{12}>\qty{80}{\percent}$, again consistent with the limit imposed by the individual source performances (Supplementary Information).

The remaining gap is therefore not set by device-to-device reproducibility but by single-source photon indistinguishability,  limited by phonon-induced pure dephasing and sideband emission for devices with $n_\text{top}=18$~\cite{Grange2017}. Using the best values reported above for $n_\text{top}=20$, and assuming perfect temporal overlap, future experiments are expected to achieve mutual indistinguishabilities approaching $M_{12}=\qty{98}{\percent}$.

%=================================================
\section*{Conclusions and perspectives}
%=================================================

In summary, we have demonstrated foundry compatible  III--V quantum-dot platform gathering the light--matter interface capabilities required for scalable hybrid photonic quantum computing. Across thousands of monolithic devices, the production pilot-line  provides reproducible control of cavity and spin parameters governing extraction efficiency, photon indistinguishability, and inter-device compatibility. We demonstrated
high-efficiency single-photon generation and collection, minute-scale photon-coherence stability, deterministic generation of multipartite spin--photon entanglement, microsecond spin coherence at low magnetic field, and high indistinguishability between photons emitted by independent sources.
These capabilities are particularly well matched to tailored spin--optical schemes for fault-tolerant quantum computation~\cite{gliniasty_spin-optical_2024,dessertaine_enhanced_2026}, in which the near one-to-one mapping between spin and photonic qubits makes the photonic layer an integral part of the computational architecture, rather than only a communication layer between matter qubits. Our platform
 also enables more efficient resource-state generation in spin--optical fusion-based schemes \cite{wein_minimizing_2025,chan_practical_2025,chen2026fusion}.

Using the current benchmarks from spin--optical fault-tolerance analyses~\cite{gliniasty_spin-optical_2024, dessertaine_enhanced_2026}, our remote-source measurements show that inter-device indistinguishability is limited by the individual source coherence, implying that the 2.3\% indistinguishability-error threshold can be reached using the single-source indistinguishabilities already achieved in optimized devices.

Likewise, the present pilot-line process, combined with improved layer doping, approaches the 6.4\% optical-loss threshold, and the proposed distributed Bragg reflector engineering is projected to bring device losses below this value.
Finally, the observed microsecond-scale spin coherence is close to the regime required for repeat-until-success gate times around 50~ns in fault-tolerant schemes~\cite{gliniasty_spin-optical_2024, dessertaine_enhanced_2026}, defining a concrete target for ultrafast control electronics and on-chip reconfiguration.

Taken together, these results shift the challenge from demonstrating the viability of the quantum-dot spin--photon interface to optimizing and integrating it at system scale for fault-tolerant quantum computing.

\section*{Acknowledgements}
This work was partially supported by the Paris Ile-de-France Région in the framework of DIM QUANTIP, the French Defense ministry - Agence de l'innovation de défense, the European Union’s Horizon CL4 program under the grant agreement 101135288 for EPIQUE project, the Plan France 2030 through the projects ANR22-PETQ-0011, ANR-22-PETQ-0006, and ANR-22-PETQ-0013. The present work was partly supported by the PROQCIMA program within the French National Quantum Strategy (France 2030), the RENATECH network, the General Council of Essonne, and EU Horizon 2020-funded QuantERA II project EQSOTIC (GA No. 101017733). Quandela thank the continual support of their investors, partners and government agencies. 
This work was also supported by the European Research Council (ERC-CoG “UNITY,” Grant No. 865230) and by the Independent Research Fund Denmark (Grant No. DFF-9041-00046B). {A.A. and K.T.L. acknowledge the National Research Foundation, Singapore through the National Quantum Office, hosted in A*STAR, under its Centre for Quantum Technologies Funding Initiative (S24Q2d0009), and OECQ through BPIFrance. K.D.J. acknowledges financial support from the BMFTR under grant FKZ I3N17238 (TUF-ToPiQC).}

\section*{Author contributions} 
H.H., H.L., T.P., P.St., and L.C. performed the advanced measurements presented in Figs. 2–4. A.Be., P.Stp., W.H., V.G., A.P., D.S., D.D.T., T.H.N.N., T.P.D., J.A.S., T.B., T.A.D.S., D.St., B.S., S.B., and T.H.A. contributed to the technological device developments. A.A., P.Stp., W.H., N.M., and J.V. performed device characterization. R.S., S.T.M., K.T.L., M.A.J., A.Au., M.Ma., S.C.W., and S.B. developed the theoretical and numerical models. S.H., Y.P., M.P., M.D., K.D.J., R.T., A.L., M.M., O.K., L.L., N.G., S.M., J.S., T.V., V.V., D.A.F., and P.S. provided experimental and scientific support. P.St., N.G., A.Au., M.Ma., S.M., J.S., T.V., V.V., S.C.W., D.A.F., S.B., and T.H.A. supervised parts of the project, while P.S. supervised the overall project. H.H., H.L., T.P., P.St., N.G., S.M., J.S., T.V., S.C.W., S.B., T.H.A., and P.S. wrote the manuscript. All authors discussed the results and commented on the manuscript.

\section*{Data and materials availability} All data that support the findings of this study are included within the article (and any supplementary files).

\section*{Competing interest} Pascale Senellart is a scientific advisor and co-founder of the company Quandela. The other authors declare no competing interests.

\bibliography{bibliography}

\end{document}